\documentclass[submission,copyright,creativecommons]{eptcs}
\providecommand{\event}{MARS 2024} 

\usepackage[frozencache,cachedir=.]{minted}

\usepackage{iftex}

\ifpdf
  \usepackage{underscore}         
  \usepackage[T1]{fontenc}        
\else
  \usepackage{breakurl}           
\fi

\usepackage{comment} 
\usepackage{graphicx}
\usepackage{subcaption}
\usepackage{xcolor}
\usepackage[final]{listings}
\usepackage[skins,listings]{tcolorbox}

\newminted{Java}{frame=lines, framerule=1pt,breaklines,autogobble}
\usepackage[normalem]{ulem}
\usepackage{float}
\usepackage{color}
\usepackage{wrapfig}
\usepackage{amssymb}
\usepackage{enumitem}
\newcommand{\cmark}{\ding{51}}%
\newcommand{\xmark}{\ding{55}}%
\usepackage{tikz}
\usetikzlibrary{spy}

\definecolor{dkgreen}{rgb}{0,0.6,0}
\definecolor{gray}{rgb}{0.5,0.5,0.5}
\definecolor{mauve}{rgb}{0.58,0,0.82}
\definecolor{navyblue}{rgb}{0.0, 0.0, 0.5}
\definecolor{ballblue}{rgb}{0.13, 0.67, 0.8}

\usepackage{algorithm}
\usepackage{algpseudocode}
\usepackage{comment}
\usepackage{amssymb}
\usepackage{pifont}
\algdef{SE}[DOWHILE]{Do}{doWhile}{\algorithmicdo}[1]{\algorithmicwhile\ #1}%
\makeatletter
\NewDocumentCommand{\LeftComment}{s m}{%
  \Statex \IfBooleanF{#1}{\hspace*{\ALG@thistlm}}\(\triangleright\) #2}

\renewcommand{\Function}[2]{%
  \csname ALG@cmd@\ALG@L @Function\endcsname{#1}{#2}%
  \def\jayden@currentfunction{#1}%
}

\newcommand{\funclabel}[1]{%
  \@bsphack
  \protected@write\@auxout{}{%
    \string\newlabel{#1}{{\jayden@currentfunction}{\thepage}}%
  }%
  \@esphack
}

\algnewcommand\algorithmicnot{\textbf{not}}
\algdef{SE}[IF]{IfNot}{EndIf}[1]{\algorithmicif\ \algorithmicnot\ #1\ \algorithmicthen}{\algorithmicend\ \algorithmicif}%

\title{
Formal Verification of Consistency
for Systems with Redundant Controllers
}

\author{
Bjarne Johansson
    \institute{ABB AB, Västerås, Sweden}
    \institute{Mälardalen University, Västerås, Sweden}
    \email{bjarne.johansson@se.abb.com}
    \and
    Bahman Pourvatan
    \quad
    Zahra Moezkarimi
    \quad
    Alessandro Papadopoulos 
    \quad
    Marjan Sirjani
    \institute{Mälardalen University, Västerås, Sweden}
\email{firstname.lastname@mdu.se}
}
\def\titlerunning{
Formal Verification of Consistency 
in Systems with Redundant Controllers}
\def\authorrunning{Bjarne Johansson et al.}
\begin{document}
\maketitle

\begin{abstract}
A potential problem that may arise in the domain of distributed control systems is the existence of more than one primary controller in redundancy plans that may lead to inconsistency. An algorithm called NRP FD 
is proposed to solve this issue by prioritizing consistency over availability.
In this paper, we demonstrate how by using modeling and formal verification, we discovered an issue in NRP FD where we may have two primary controllers at the same time.
We then provide a solution to mitigate the identified issue, thereby enhancing the robustness and reliability of such systems. 
\end{abstract}

\section{Introduction}\label{sec::Introduction}
Control systems are essential in the automation solution of domains such as offshore oil extraction, refineries, and hydropower plants - sectors where downtime can lead to significant financial losses  or even life-threatening incidents. These automation solutions incorporate redundancy to mitigate the risk of unplanned downtime due to hardware failures by duplicating critical components like controllers.  The common approach is standby redundancy, where an active primary controller manages the process, and a passive backup is ready to take over in case of primary failure~\cite{Simion2023PLCRedReview}. 
These controllers, or Distributed Controller Nodes (DCN), interact with the physical world through Field Communication Interfaces (FCI), connecting to input/output (I/O) devices. The FCI supplies process values to the DCN, which then executes control actions based on these inputs and sends outputs back to the FCI. 
\begin{wrapfigure}{r}{0.35\textwidth}
	\centering
\includegraphics[scale=0.90, bb=0 0 290 120]{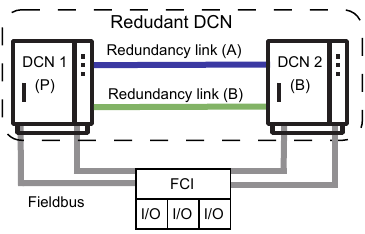}
    \caption{A redundant DCN (controller) pair synchronized with dedicated, redundant redundancy link.} \label{fig:SimpRedDCN} 
\end{wrapfigure} 
For a backup DCN to seamlessly assume the primary role, it must detect the primary's failure and resume the primary role with the former primary's last known state. The primary cyclically replicates its latest state to the backup and sends a heartbeat, i.e., a message with predetermined intervals for failure detection. Heartbeat absence signifies a possible primary failure. Controller redundancy communication is conventionally carried out over a dedicated, point-to-point connection~\cite{Stoj2020CostEfcRedPlc,Emerson2022CpuRed,Siemens2023RedSys}, as illustrated in Figure~\ref{fig:SimpRedDCN}. Failure of the redundancy link can partition the DCN pair, disrupting synchronization and causing their internal states to diverge. This divergence might result in inconsistent outputs to the FCI.

Two strategies are common when managing failures in redundancy communication links: (i) disabling redundancy following the failure of one of the links or (ii) continuing in redundant mode. These strategies reflect the alternatives a distributed system has in case of partitioning: remain consistent and sacrifice availability  
or vice versa—consequence of the Consistency, Availability, and Partitioning tolerance (CAP) theorem~\cite{Gilbert2002CAP}.

Disabling redundancy after a redundancy link failure compromises availability, as the backup won't activate if the primary controller fails before the link is repaired. While this method prioritizes consistency, a concurrent loss of both redundancy links can still lead to a dual primary situation~\cite{Siemens2023RedSys}.

The alternative, operating redundantly with only one functioning redundancy link, risks causing a dual primary situation if the remaining redundancy link fails.  
This is because the backup can not distinguish missing heartbeats due to a failure of the link from a failure of the primary. Some vendors call a dual primary scenario non-synchronized active units, signifying the consistency compromise following from CAP~\cite{Emerson2022CpuRed}. Controllers unable to communicate can not synchronize, leading to an inconsistent state in the redundant pair.

\begin{figure}[htb!]
	\centering
 \includegraphics[scale=0.80, bb=0 0 290 110]{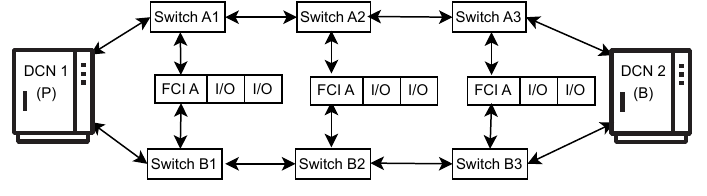}
    \caption{Redundant controllers connected over a redundant, disjoint network backbone.} \label{fig:NetwRedDCN}  
\end{figure}

The advent of Industry 4.0 is steering industrial controllers towards a network-centric design~\cite{ABB2022DCSVision,Akerberg2021FutureIndNetw,Leander2023NCC}. As defined by the Open Process Automation Forum (OPAF), the DCNs and FCI are integrated into a cohesive communication network. Additionally, this network backbone can support redundancy communication and replace the redundancy link shown in Figure~\ref{fig:SimpRedDCN} with a network, see Figure~\ref{fig:NetwRedDCN}.

\begin{figure}
\centering
 \begin{subfigure}[htb!] {0.48\textwidth}
	\centering
	\includegraphics[scale=0.65, bb=0 0 320 130]{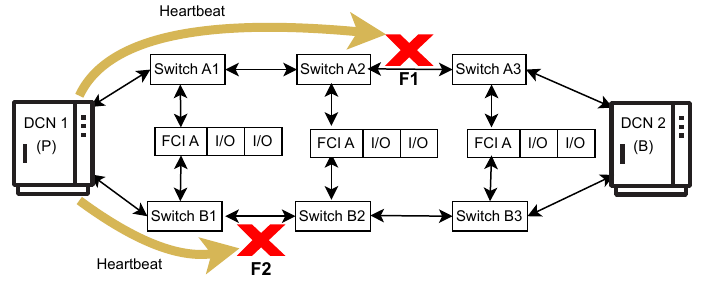}
    \caption{}
    \label{fig:NetwRedFail} 
    \end{subfigure}  
    \hfill
\begin{subfigure}[htb!]{0.48\textwidth}
	\centering
	\includegraphics[scale=0.65, bb=0 0 320 130] 
{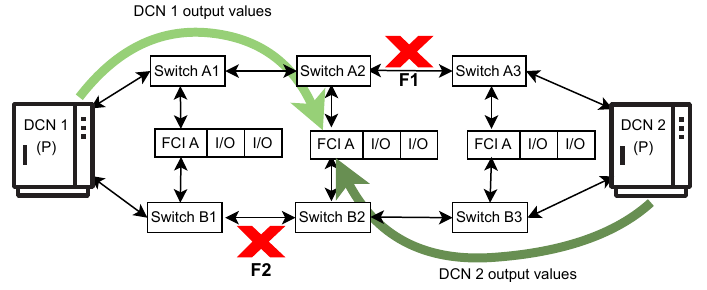}
  \caption{}
  \label{fig:NetwRedFailIncon}  
\end{subfigure} 
\caption{(a) F1 and F2 exemplify network failures partitioning the redundant controller pair, preventing the heartbeat (and other communication) between DCN 1 and DCN 2. (b) Due to F1 and F2 caused partitioning, both DCN 1 and DCN 2 become primary and drive potentially inconsistent outputs.} 
\end{figure}

When communication between a redundant DCN pair fails, as shown in Figure~\ref{fig:NetwRedFail}, traditional approaches either disable redundancy at the first failure (F1) or allow the system to operate in a non-synchronized dual-primary mode, as shown in Figure~\ref{fig:NetwRedFailIncon}. Johansson et al.~\cite{Johansson2023NRPFD} introduce the Network Reference Point Failure Detection (NRP FD) for such redundant DCN systems. NRP FD prioritizes consistency while reducing the impact on availability. It uses an external Network Reference Point (NRP) as a tiebreaker for primary role determination, aiding the backup DCN in differentiating between primary and network failures. For a DCN to attain and retain the primary role, it must maintain communication with the NRP. The importance of addressing dual primary risks is emphasized in manuals recommending spatially separated redundancy links in current systems to avoid simultaneous damage and undefined system states~\cite{Siemens2023RedSys}.

To design an algorithm that guarantees the uniqueness of the  primary the following questions need to be answered:
\begin{itemize} \item How should the backup know about a failure?  \item When should the backup become a primary?\end{itemize}

 \noindent As described by Johansson et al. in~\cite{Johansson2023NRPFD}, the NRP FD uses heartbeats for primary failure detection (heartbeat) and a separate message for NRP reachability testing and detecting network failure.
This introduces a potential vulnerability: the absence of a heartbeat is a sign of the primary failure, while NRP reachability is verified separately.
Consequently, temporary disturbances could lead to inconsistencies, underscoring the importance of testing with temporal disturbances. Hence, one other question also have to be answered:
\begin{itemize} 
\item 
    How should we take care of the transient errors in switches or DCNs?
\end{itemize}

Since nondeterministic behavior is generally undesirable in control systems, particularly in high-integrity systems crucial for safety-critical solutions, like the ABB AC 800M High Integrity system \cite{ABB2024HiDCS}, we need assurance of the correctness of the algorithm. 
Therefore, this paper describes in detail the modeling and formal verification of the NRP FD algorithm, considering the main safety property of "NoDualPrimary". 
We use Timed Rebeca which is an actor-based modeling language for reactive and distributed systems and its model checker tool Afra to model and verify NRF FD. 
We model different failures including transient errors and illustrate the results. We also propose an  enhanced lease-based version of NRP FD that ensures a singular primary in the case of transient errors.

\section{Network Reference Point Failure Detection (NRP FD) Algorithm}\label{sec::NRP-FD}
\begin{figure}[htb!]
	\centering
	\includegraphics[scale=0.80, bb=0 0 310 130]{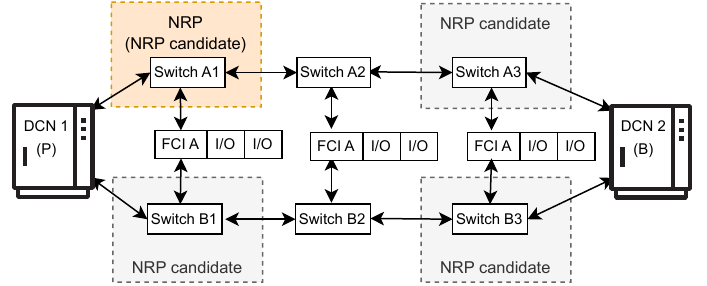}
    \caption{The redundant network backbone with the NRP and NRP candidates highlighted.} \label{fig:NRP-Top}  
\end{figure}

NRP FD targets failure detection in redundant controller pairs. In a standard system, two controllers, DCN 1 and DCN 2, function as primary and backup, respectively, as illustrated in Figure~\ref{fig:NRP-Top}. The primary is unique in the system and interacts with I/O devices, while the backup, in standby mode, activates only upon primary failure. This concept is known as standby redundancy~\cite{Simion2023PLCRedReview}. These controllers, DCN 1 and DCN 2, require communication, typically through a network facilitated by switches~\cite{Akerberg2021FutureIndNetw,Leander2023NCC}. Redundant controllers are often paired with dual independent networks for enhanced reliability, as depicted in Figure~\ref{fig:NRP-Top}.

NRP FD is a heartbeat-based failure detection algorithm where the primary controller sends regular heartbeat messages to the backup via the networks connecting the redundant DCN pair \cite{Johansson2023NRPFD}. 
These heartbeats, a push-based failure detection method, involve the primary sending messages to the backup at a known interval~\cite{Satzger2007NewFailureDetector}. NRP FD differs from traditional heartbeat-based failure detection due to its NRP usage. An NRP must meet two requirements: (i) it should not share common cause failures with the redundant DCN pair, and (ii) be accessible from only one DCN in case of network partitioning. Each controller typically has one NRP candidate per independent network, as illustrated in Figure~\ref{fig:NRP-Top}, where network switches serve as potential NRPs. The uper network in Figure~\ref{fig:NRP-Top} includes three switches $Switch~A1$, $Switch~A2$, and $Switch~A3$, and the lower network includes  $Switch~B1$, $Switch~B2$, and $Switch~B3$. The NRP candidate set for the primary is $\{Switch~A1, Switch~B1\}$ and for the backup is $\{Switch~A3, Switch~B3\}$, and $Switch~A1$ is the NRP.

The operational procedure of NRP FD is as follows: before enabling redundancy, the primary DCN selects an NRP from the available NRP candidates. 
The heartbeat message communicates the NRP selection to the backup. The primary continuously monitors the NRP, ensuring its accessibility and proposing a change to the backup if the NRP is unreachable. If the backup doesn't acknowledge this change within a set time, the primary leaves the primary role.
Concurrently, the backup continuously monitors heartbeats from the primary. If these are missing for a predetermined duration, the backup assesses its NRP connection. Should this connection be active, the backup takes the primary role.
The following section will provide more details of the algorithm and its Timed Rebecca model.

\section{Modeling and Verification of NRP FD using Timed Rebeca}\label{sec::Modeling}

We use Timed Rebeca language and its integrated model checker tool, Afra, 
to model and verify NRP FD.  
For modeling NRP FD, we have used the description of the protocol and the diagrams provided in \cite{Johansson2023NRPFD} as well as several meetings with the industrial partners to clarify the details and choose the appropriate level of abstraction, which we will discuss in the remainder of this section.  

\subsection{The actor-based language, Timed Rebeca}\label{subsec::Rebeca}
Rebeca (Reactive Object Language) \cite{Sirjani-Rebeca,Rebeca-Sirjani06} is an actor-based language designed for  modeling and formal verification of reactive concurrent and distributed systems. 
Actors \cite{hewitt1973universal,Agha-Actors-1986} are units of concurrency. 
In Rebeca models, reactive objects known as rebecs resemble actors with no shared variables, asynchronous message passing, and unbounded message buffers.
Each rebec has a single thread of execution. 
Communication with other rebecs is achieved by sending messages, and periodic behavior is executed by sending messages to itself.
Rebeca has no explicit receive statement, and its send statements are non-blocking. Each rebec has variables, methods (message servers), and a dedicated message queue for received messages.
How a rebec reacts to a message is specified in message servers.
The rebec processes messages by de-queuing from the top and executing the corresponding message server non-preemptively. 
The state of a rebec can change during the execution of its message servers through assignment statements.

Rebeca is an imperative language with a syntax similar to Java. A Rebeca model consists of several reactive classes and a main section. Each reactive class describes the type of a certain number of rebecs.  Rebecs (actors) are instantiated in the main block. 
While message queues in the semantics of Rebeca are inherently unbounded, a user-specified upper bound for the queue size is necessary to ensure a finite state space during model checking. 
Reactive classes include constructors, sharing the same name as the class, responsible for initializing the actor's state variables and placing initially required messages in the actor's message buffer.

In this work, we use Timed Rebeca (the timed  extension of Rebeca) \cite{Sirjani2016TimeActors,Khamespanah-shiftEq-2015}   
with a global logical time. Timed Rebeca considers synchronized local clocks for all actors throughout the model.   
Instead of a message queue, Timed Rebeca uses a message bag in which messages carry their respective time tags. 
The sender tags its local time to a message at the time of sending. 
Timed Rebeca introduces three timing primitives: "delay," "after," and "deadline." 
A delay statement represents the passage of time for an actor while executing a message server, i.e., it is used to model computation times. All other statements are assumed to execute instantaneously.
The keywords "after" and "deadline" are augmented to a message send statement. The term "after(n)" means it takes n units of time for a message to reach its receiver. 
Using the after construct, we can model network delay and periodic events. We can use a nondeterministic assignment to n, and model nondeterministic arrival times for a message (event).
The term "deadline(n)" conveys that if the message is not retrieved within n units of time, there will be a timeout.
An abstract syntax of Timed Rebeca is provided in Appendix \ref{app::AbstractSyntaxofTimedRebeca}. Timed Rebeca is extended with priorities \cite{Sirjani-Priority-2020}. Priorities 
are assigned to rebecs and message handlers to control the order of their execution and hence  enhance the determinism of the system's behavior \cite{Khosravi-ActorsUpgraded-2023}. 
If more than one actor or event are enabled at the same time, then the model checker builds all the possible execution traces, using priorities you can cut some of the branches.

\subsection{Modeling NRP-FD in Timed Rebeca} \label{subsec::Modeling}
We model Figure \ref{fig:NRP-Top} using Timed Rebeca. The model is extensible meaning that the number of switches and nodes can be increased.
In the Timed Rebeca model each node and each switch is modeled as an actor, their communication is modeled as message passing, and reactions to each message, signal, and timed event are modeled using message servers.
A Rebeca model includes reactive class definitions, defining the behavior of the rebecs (actors) within the model. 
L\ref{Rebeca}\footnote{We use L1, L2 and L3 to refer to Listing~\ref{Rebeca}, Listing~\ref{BACKUP-Code} and Listing~\ref{Property}, respectively.} illustrates some parts of the Timed Rebeca model for NRP FD.

\begin{listing}[htp!]
\scriptsize
\begin{minted}[frame=lines,framesep=1mm,xleftmargin=4em,linenos,escapeinside='']{java}
env int heartbeat_period = 1000;
env int max_missed_heartbeats = 2;
env int ping_timeout =500;  
env int nrp_timeout = 500;
env byte NumberOfNetworks = 2;
env int switchA1failtime = 2500; 
...
env int networkDelay = 1;
env int networkDelayForNRPPing = 1;
reactiveclass Node (4){ 
    knownrebecs {Switch out1, out2;}
    statevars {...}
    Node (int Myid, int Myprimary, int NRPCan1_id, int NRPCan2_id, int myFailTime) {
        id = Myid;
        NRPCandidates[0] =NRPCan1_id;
        NRPCandidates[1] =NRPCan2_id;	
        NRP_network = -1;
        primary = Myprimary;
        mode = WAITING;
        ...
        if(myFailTime!=0) nodeFail() after(myFailTime);
        runMe();        
    }
    msgsrv new_NRP_request_timed_out(){...}
    msgsrv ping_timed_out() {...}  
    msgsrv pingNRP_response(int mid){...} 
    msgsrv new_NRP(int mid,int prim, int mNRP_network, int mNRP_switch_id) {...}
    msgsrv runMe(){ 
        if(?(true,false)) nodeFail();
        switch(mode){
            case 0: //WAITING : ...
            case 1: //PRIMARY : ...
            case 2: //BACKUP : ...
            case 3: //FAILED : ... 
        self.runMe() after(heartbeat_period);
     }
     msgsrv heartBeat(byte networkId, int senderid) {...}
     msgsrv nodeFail(){...}
}
reactiveclass Switch(10){
    knownrebecs {...}
    statevars {...}
    Switch (int myid, byte networkId, boolean endSwitch , Switch sw1, Switch sw2, int myFailTime) {
       mynetworkId = networkId;
       id = myid;
       terminal=endSwitch;
       amINRP = false;
       failed = false;
       switchTarget1 = sw1;
       switchTarget2 = sw2;
       ...
    }
    msgsrv switchFail(){ failed = true; amINRP=false;} 
    msgsrv pingNRP_response(int senderNode){...}
    msgsrv pingNRP(int switchNode, int senderNode, int NRP) {...}
    msgsrv new_NRP(int senderNode, int mNRP_network, int mNRP_switch_id) {...}
    msgsrv heartBeat(byte networkId, int senderNode) {...}
}
main {
    @Priority(1) Switch switchA1(DCN1):(1, 0, true , switchA2 , switchA2 , switchA1failtime);
    @Priority(1) Switch switchA2(DCN1):(2, 0, false , switchA1 , switchA3 , switchA1failtime);
    @Priority(1) Switch switchA3(DCN2):(3, 0, true , switchA2 , switchA2 , switchA3failtime);
    @Priority(1) Switch switchB1(DCN1):(4, 1, true , switchB2 , switchB2 , switchB1failtime);
    @Priority(1) Switch switchB2(DCN1):(5, 1, false , switchB1 , switchB3 , switchB1failtime);
    @Priority(1) Switch switchB3(DCN2):(6, 1, true , switchB2 , switchB2 , switchB3failtime);
    @Priority(2) Node DCN1(switchA1, switchB1):(100, 100, 1, 4, node1failtime);
    @Priority(2) Node DCN2(switchA3, switchB3):(101, 100, 3, 6, node2failtime);
}
\end{minted}
\caption{(L1) An abstracted version of the Timed Rebeca model of NRP FD (Full version in Appendix~\ref{app::TimedRebecaNRPFD}).}
\label{Rebeca}
\end{listing}

In the NRP-FD model, we have two different element types, Node and Switch. Each element type is defined as a reactive class,  $Node$ (L\ref{Rebeca}, line 10) and $Switch$ (L\ref{Rebeca}, line 40). 
Each reactive class has a constructor. A constructor is a unique method which is called when the actor is instantiated. 
Initialization of the variables is done in the constructor. 
We instantiate two nodes with ids 100 and 101 and six switches (A1-A3 and B1-B3) in the main section (L\ref{Rebeca}, lines 59-68). A node can be a primary or a backup, and a switch can be a non-terminal switch (not connected to a DCN), an NRP candidate, or an NRP. 
Each node has an NRP candidate (switch) for each network, i.e., switches A1 and B1 with ids 1 and 4, respectively for DCN1 and switches A3 and B3 for DCN2 with ids 3 and 6, respectively (L\ref{Rebeca}, lines 66-67). 
The parameters in the instantiation statements are used to set different types and also pass other necessary information to the constructor.

We select DCN1 with id 100 as the primary at the beginning of the algorithm (second parameter in lines 66-67 of L\ref{Rebeca}).
There are two known rebecs in the reactive class $Node$, meaning it can  
send messages to these rebecs. 
We have a method call in the constructor of the $Node$, i.e., $runMe$ (L\ref{Rebeca}, line 22). In $runMe$ (L\ref{Rebeca}, line 28)  
the DCN checks its state  using the state variable $mode$ and then serves the corresponding behavior (L\ref{Rebeca}, lines 30-34).
Note that, the last line of $runMe$ (L\ref{Rebeca}, line 35) is a self-call followed by an after with $heartbeat\_period$ as its parameter, modeling a periodic event, i.e., "$runMe() after(heartbeat\_period);$". It means that in every $heartbeat\_period$ (determined in the code L\ref{Rebeca}, line 1), $runMe$ is executed.
The $heartbeat\_period$ should be significantly larger than other timing parameters. This is  because all events must be handled during a heartbeat interval. 
Regarding timing parameters in modeling, we carefully consider values so that the model matches the reality.
We will discuss more on timing in the following.

\begin{figure}[htb!]
	\centering
    \includegraphics[scale=0.70, bb=0 0 430 190]{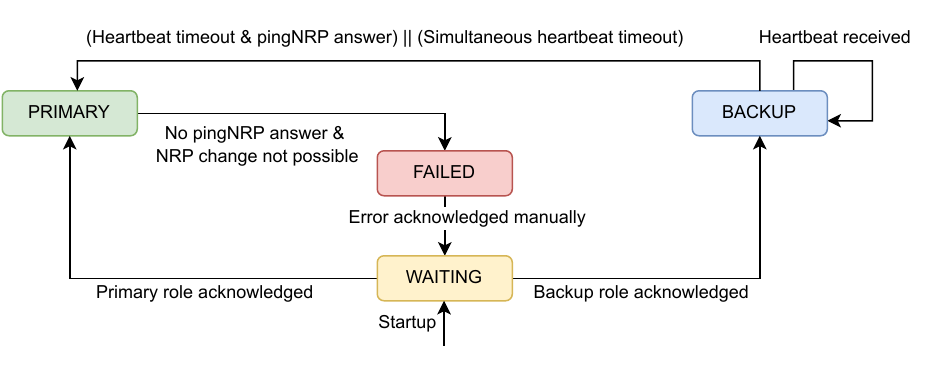}
    \caption{
    Different modes of a DCN in NRP FD in the Rebeca model.
$WAITING$ is the initial mode. The node transitions from $WAITING$ to $PRIMARY$ or $BACKUP$ based on the value passed to its constructor.
From $PRIMARY$, it moves to $FAILED$ if after sending a $pingNRP$ it receives no response from NRP within the deadline,  and it cannot change the NRP either. In the $BACKUP$ state, the node transitions to $PRIMARY$ if the heartbeat timeouts and $pingNRP$ detects a responsive NRP, or when the heartbeat timeout occurs simultaneously for both networks. In the latter case the backup node assumes that the primary node failed, because it is unlikely that there is a failure in both networks.
The node stays in $BACKUP$ mode as long as it is receiving heartbeats.
It remains in $FAILED$ until the situation is resolved manually.} \label{fig:NrpFdStates}  
\end{figure}

In NRP FD, DCNs have four modes, $WAITING$, $BACKUP$, $PRIMARY$, and $FAILED$, as detailed in the diagram in Figure~\ref{fig:NrpFdStates}. 
In the Rebeca model, we 
set the initial mode of DCN to $WAITING$ in the constructor of $Node$, L\ref{Rebeca}, line 19.
We pass the primary id to both nodes and in the $WAITING$ mode the variable denoting the role is set accordingly, and an NRP is announced.
 
In the $PRIMARY$ mode, the primary DCN 
tests the NRP reachability with $pingNRP$, i.e., sends message $pingNRP$ to the NRP which then is served using the message server $pingNRP$ (L\ref{Rebeca}, line 55). In a real system, the $pingNRP$ could be realized with an Internet Control Message Protocol (ICMP) echo (commonly known as ping) or another suitable protocol depending on the NRP's capabilities. If the NRP fails to respond, the primary announce a new NRP, assuming alternatives are available (using $new\_NRP$ message server, L\ref{Rebeca}, line 56).
After assuring that an NRP exists, the primary DCN sends heartbeats.
If there is no available NRP, the primary transition to the $FAILED$ mode ($ping\_timed\_out$ in L\ref{Rebeca}, line 25).

In the $BACKUP$ mode, the DCN expects heartbeats from the primary. The heartbeat period and tolerance limits (i.e., the number of missed heartbeats before a timeout is declared) must be carefully set to minimize false positives due to transient disturbances. Given that typical DCN redundancy involves two disjoint network paths, a heartbeat is expected on each network path per period. Simultaneous timeouts on all paths likely indicate a primary failure rather than failure of both networks. Thus, NRP FD offers an optimization: transitioning directly to the $PRIMARY$ mode upon simultaneous heartbeat timeouts, bypassing the $pingNRP$ exchange. However, this optimization slightly increases the risk of dual primaries. This is a bug that model checking catches.
The number of maximum missed heartbeats is set to 2 ($max\_missed\_heartbeat$ in L\ref{Rebeca}, line 2). 
L\ref{BACKUP-Code}, shows the $BACKUP$ part of the message server $runMe$. 
The variables $heartbeats\_missed\_1$ and $heartbeats\_missed\_2$ are counters for heartbeats on the two networks which will increase at each period, and is reset to zero when a heartbeat is received.
The backup DCN counts consecutive $heartbeats\_missed$ for each network. 
If both counters exceed the defined limit of $max\_missed\_heartbeat$ 
(L\ref{BACKUP-Code}, line 4), the backup detects a failure and sends a $pingNRP$  to the NRP to verify its reachability. If the NRP is reachable, the DCN  
transitions from $BACKUP$ to the $PRIMARY$ state (in $ping\_timed\_out$, L\ref{Rebeca}, line 25).

In the $FAILED$ mode, NRP FD awaits the acknowledgment that manually confirms the resolution of the issues that triggered the transition to 
$FAILED$. 
\begin{listing}[h]
\scriptsize
\begin{minted}[frame=lines,framesep=1mm,xleftmargin=2em,linenos,escapeinside='',breaklines]{java}
case 2: //BACKUP : 
 heartbeats_missed_1++;
 heartbeats_missed_2++;
 if (heartbeats_missed_1 > max_missed_heartbeats && heartbeats_missed_2 > max_missed_heartbeats){
  if(heartbeats_missed_1==heartbeats_missed_2 && heartbeats_missed_2==max_missed_heartbeats+1){
   mode = PRIMARY;
   primary=id;
   ...
  }else{
   heartbeats_missed_1 = (heartbeats_missed_1>max_missed_heartbeats+2)?max_missed_heartbeats+2:heartbeats_missed_1;
   heartbeats_missed_2 = (heartbeats_missed_2>max_missed_heartbeats+2)?max_missed_heartbeats+2:heartbeats_missed_2;
   if(NRP_network==0){	
    ping_pending = true;
    NRP_network=-1;
    out1.pingNRP(id, NRP_switch_id) after(5);
    ping_timed_out() after(ping_timeout);
   }else{ ...  // the other network }
   NRP_pending = true;
  }
 }
 else if(heartbeats_missed_1 > max_missed_heartbeats || heartbeats_missed_2 > max_missed_heartbeats){...}
\end{minted}
\caption{(L2) The behavior of a DCN in $BACKUP$ mode, in the message server $runMe$ (full version is provided in Appendix~\ref{app::TimedRebecaNRPFD}).
}
\label{BACKUP-Code}
\end{listing}

\textbf{Accuracy of the model.} Based on the real situation, we consider the topology and the way the DCNs interact with each other. 
The rationale for tolerating up to two lost heartbeats ($max\_missed\_heartbeats = 2$) is based on the low bit error rate of gigabit Ethernet and the ability of a heartbeat message to fit within a standard 1500-byte Ethernet frame. This suggests a low likelihood of losing heartbeat messages, especially across two disjoint networks, thus minimizing the risk of false positives due to regular disturbances.
The $heartbeat\_period$, combined with $max\_missed\_heartbeats$, determines the reaction time - 
the duration from the occurrence of a primary failure to the point at which the backup takes over the primary role.
The takeover time requirement varies by domain; for process control, a maximum of 500 milliseconds is tolerable, as suggested by Hegazy et al.~\cite{Hegazy2015AutomationAsCloudService}. System manuals indicate feasible heartbeat periods are in the tens of milliseconds range\cite{Siemens2023RedSys,Emerson2022CpuRed}.
Regarding propagation and $pingNRP$ response times, the propagation of a full-sized Ethernet frame on Gigabit Ethernet is about 12 microseconds, negligible compared to the heartbeat period. The NRP's response time is implementation-dependent, potentially under a millisecond. If ICMP ping is employed, a few milliseconds response times are achievable~\cite{Johansson2023NRPFD}.
We've defined the $heartbeat\_period$ as 1000 time units and set the $ping\_timeout$ and $nrp\_timeout$ to 500 time units. We also consider $networkDelay$ and $networkDelayForNRPPing$ as 1 unit of time.  We use the keyword $after$ when DCNs  ping the NRP node and set it to 5 units of time. 
These values are chosen to be approximately close to the actual values and preserve the sequence of the messages.
Therefore, they may vary, for instance, to a greater or lesser extent. But all timing events should be handled within one period, 1000 time units in our model.
We used the $after$ construct where we needed to respect the order of execution.

\section{Model checking of NRP FD using Afra}

We can define our desired properties using assertions in a separate file in Afra and perform model checking. A snapshot of Afra is provided in Appendix \ref{app::Afra}. The main safety property, "$NoDualPrimary$," is shown in L\ref{Property}. This property is set to recognize the dual primary state, i.e., in no state the modes of the two DCNs are both primary.
We first define a set of atomic propositions, and then the assertions based on these propositions.
Timed Rebeca has a TCTL model checking but it is not integrated in Afra.
In many cases, looking at the visualization of the state space helps us see the problems with the algorithm.

\begin{listing}[htp!]
\scriptsize
\begin{minted}[frame=lines,framesep=1mm,xleftmargin=4em,linenos,escapeinside='']{java}
property { 
 	define {
 		DCN1Primary = (DCN1.mode ==1);
 		DCN2Primary = (DCN2.mode ==1);
 	}
Assertion{ NoDualPrimary:!(DCN1Primary && DCN2Primary); }}
\end{minted}
\caption{The safety property "$NoDualPrimary$" for NRP FD.}
\label{Property}
\end{listing}

\noindent For model checking, we consider the regular system behaviour, and scenarios where we have failures of DCNs and switches. We examine all the possible failure combinations of DCNs and switches at the start of handling an event, and perform model checking to provide a comprehensive analysis.
We have modeled failures in three scenarios each of which can have different cases: \\
   \noindent \textbf{1. Failures on each event.}
   In this scenario, we add the following commands at the beginning of each message server for DCNs and switches, simulating the possibility of their failure. 
    Since this scenario models the failure where an event should be handled, we refer to it as event-based. The expression "$?(true,false)$" represents a nondeterministic choice between true and false. When the value true is chosen then a variable is set, this variable is checked in the beginning of  the messages servers and if it is set the message server is not executed. 
     
    \vspace{-2mm}
    \lstset{language=Java}
    \begin{lstlisting}[belowskip=-0.2 \baselineskip]
    //Possible failure for a DCN: 
    if(?(true,false)) nodeFail();
    //Possible failure for a Switch: 
    if(?(true,false)) switchFail();
    \end{lstlisting}
    \vspace{2mm}
    \noindent \textbf{2. Failures that occur at specific times.} We define a set of variables to model the failure of different DCNs and switches at specific times. By manipulating these variables, we can model various combinations of DCN and switch failures at different times across multiple model checking runs. 
    \vspace{-2mm}
    \lstset{language=Java}
    \begin{lstlisting}[belowskip=-0.2 \baselineskip]
    env int switch1failtime = 0; env int switch2failtime = 2500; env int switch3failtime = 0;
    env int switch4failtime = 2500;
    env int node1failtime = 0; env int node2failtime = 0;
    ..
    //Failure of a DCN at a specific point of time. Value zero means no failure. 
    if(myFailTime!=0) nodeFail() after(myFailTime);
    ...
    //Failure of a Switch at a specific point of time. Value zero means no failure. 
    if (myFailTime!=0) switchFail() after(myFailTime);
    \end{lstlisting}
    \vspace{2mm}
    \noindent \textbf{3. Transient failures.} These failures could occur, for example, if an attacker deliberately drops the heartbeats for more than the maximum allowed misses ($max\_missed\_heartbeats$) on both networks. 
    Subsequently, the backup DCN, upon detecting missed heartbeats, checks the NRP. If the NRP is reachable, it becomes the primary, assuming that the primary has failed, resulting in a dual-primary situation. So we model a transient failure where both heartbeats are missed. Part (not all) of the code for this scenario is the following, which states that only if we do not have an attacker, then the heartbeats will be sent. 
     \vspace{-2mm}
    \begin{lstlisting}[belowskip=-0.2 \baselineskip]
    if(attacker<1){
        out1.heartBeat(0, id) after(networkDelay);
        out2.heartBeat(1, id) after(networkDelay);
    }
    \end{lstlisting}
   
Table~\ref{tab:configurationofFailures} illustrates the scenarios we have considered and checked. Number of states and transitions are also reported. Note that in the cases where the the assertion is violated, model checking is stopped after reaching a counter example.
Case 1 is the case with no failure. Case 2 is the event-based failure scenario where we investigate all combinations of failures for any DCN or switch, where they stop reacting to the events.
Cases 3 to 5 consider failures at time 2500 for DCN1, switchA1 and switchA3, respectively. 
This number is intended to go through a full round of algorithm execution, with two heartbeats.
We have considered case 3 for the $PRIMARY$ failure as DCN1 is initially set as the primary DCN. We also consider cases 4, 6, and 7 as failures of switches A1 and B1 can cause the primary DCN to be disconnected from the networks.
Case 5 is also considered to model a situation where the backup cannot ping the NRP. 
Case 8 is modeling the transient error. There are three cases where the model violates the property.  

\begin{table}[htp!]
  \small
  \centering
  \caption{Different test scenarios, without any failures, and with different types of failures
  }
\begin{tabular}{clcc}
Case & Configuration for failures                             & Result  & no. of states and transitions    \\ \hline
1    & Without failure                & {\color[HTML]{32CB00} \cmark} & 38, 49\\
2    & Failures on each event & {\color[HTML]{F56B00} \xmark} & 3539, 4677\\
3    & DCN1 fails at time 2500                                & {\color[HTML]{32CB00} \cmark} & 113, 138 \\
4    & switchA1 fails at time 2500                            & {\color[HTML]{32CB00} \cmark}  & 114, 134 \\
5    & switchA3 fails at time 2500                            & {\color[HTML]{32CB00} \cmark}  &  146, 179 \\
6    & switchA1 fails at time 2500 and switchB1 at time 3500  & {\color[HTML]{32CB00} \cmark}  & 187, 223 \\
7    & switchA1 and switchB1 fails simultaneously at time 2500 & {\color[HTML]{F56B00} \xmark}  & 70, 88 \\
8    & Heartbeats are missing  because of transient errors     & {\color[HTML]{F56B00} \xmark}  & 35, 42 

\end{tabular}
\label{tab:configurationofFailures}%
\end{table}
Afra generates a counter-example in cases of any violation (here for cases 2, 7 and 8 of Table~\ref{tab:configurationofFailures}). 
We can explore the states in the counter-example and see the value of the state variables in each of them. 
A snapshot of the state space showing the dual primary situation for the case 7 is depicted in Figure \ref{fig:StateSpace-NRPFD} of Appendix \ref{app::StateSpace}. These cases may be rare situations in reality, but in formal verification we detect and eliminate the corner cases. To overcome these issues, we provide an extension for NRP FD, which will be described next. 

\subsection{Leasing NRP FD}
To address failure issues, we provide an enhanced NRP FD version called Leasing NRP FD. First, we remove the optimization, i.e., transitioning directly from $BACKUP$ to the $PRIMARY$ mode upon simultaneous heartbeat timeouts, bypassing the $NRPPing$ exchange (L\ref{BACKUP-Code}, lines 5-9). 

While NRP FD prioritizes consistency, even without optimization, there remains a non-zero probability of failure.
The heartbeat and $pingNRP$ messages are separate: the heartbeat indicates whether the primary is alive, and the $pingNRP$ informs the backup about its separation from the NRP or the NRP's failure. Since these messages are distinct and can be independently disrupted, it's theoretically possible, as indicated by verification, that a temporary disturbance might disrupt the heartbeats. This disruption could lead the backup to believe the primary has failed, and upon a successful $pingNRP$ following the transient disturbance, it might erroneously become the $PRIMARY$, even while the other DCN remains primary.
To address this vulnerability, we introduce the Leasing NRP FD, where the primary role is 'leased' from the NRP. This leasing can be implemented in various ways. In our model, the NRP timestamps the latest $pingNRP$ from the primary, and then the backup checks this timestamp. 
Full version of Leasing NRP FD is provided in Appendix~\ref{app::TimedRebecaNRPFD} and also on the Rebeca GitHub page\footnote{https://github.com/rebeca-lang}. 
Even with a low probability of dual primary occurrences in the original NRP FD, this inherent algorithmic trait could lead to nondeterministic behavior, which is
unacceptable in safety-critical solutions. Thus, there's a need for algorithms like Leasing NRP FD, which eliminate such violations and are more suitable for safety-critical systems.
For this new algorithm, Afra created
15891 states, and 34053 transitions, and the assertion is satisfied.

\section{Why Timed Rebeca?}\label{subsec::WhyTimedRebeca}
In \cite{Sirjani2018Friendliness}, Sirjani argues that when selecting a modeling language, expressiveness is a key factor, but faithfulness to the system being modeled and usability for the modeler are equally crucial.
Faithfulness is about how similar the model and the system are.
It determines if and how the structures and features supported by the modeling language match with the requirements of the system's domain.
Faithfulness makes reusability possible, also in cases gives us better analyzability and traceability.
Usability concerns the modeler,  and how swiftly the modeler can use the language.  These two aspects together are called as friendliness in \cite{Sirjani2018Friendliness}.

Timed Rebeca is a language for modeling asynchronous communication in  distributed systems, incorporating a focus on time-related aspects. 
Regarding faithfulness, actors are units of concurrency like the controllers and switches in our case study.
Timed Rebeca is event-driven, taking messages/events from the message/event bags and executing their corresponding message servers. Timed Rebeca is used for modeling and verification in many domains including  different network protocols, schedulability in sensor networks and   Network on Chip (NoC)
\cite{Sirjani2018Friendliness}.
Considering our problem in the domain of distributed control systems, Timed Rebeca provides a natural mapping of
structures, features, and flow of control for our purpose such as modeling the topology of the network, behavior of the DCNs and switches based on their roles, the way they  communicate using message passing,
progress of time required for handing a message, network delay, and periodic events using primitive timing keywords.
Message queues/buffer are not explicit and the modeler does not need to manage them. Timing concept is intuitive, and you model the behavior from the perspective of each actor.

Regarding usability, it has a structure like a programming language, hence, it is easy for programmers to use. Debugging can be done based on the counterexamples and going through the  model checking process iteratively.  
Timed Rebeca is supported by an 
Eclipse IDE called Afra \cite{Khamespanah-Afra}. Afra provides a model checker tool for the family of Rebeca languages. 
The modeler enters the model and the properties in separate files, then model check and debug the model in Afra. 
Timed models result in an infinite number of states in the state space due to the progress of time, leading to unbounded transition systems. 
A shift-equivalence relation is introduced for Timed Rebeca in \cite{Khamespanah-shiftEq-2015,khamespanah2015floating} 
to ensure a bounded state space.
Afra utilizes this relation to generate the state space including local actor states and logical time. 
Desired properties can be written as assertions in a separate file in Afra. In case of violation, a counter-example is shown visually alongside the model which gives us the ability to traverse and check the values of the actors' variables. 
As the state space is provided in an XML file, 
it is also possible to have a visual representation of the entire state space (see an example in Figure \ref{fig:StateSpace-NRPFD}, App. \ref{app::StateSpace}). 
All the above gives us a natural and easy way to model our system, and also provide us  analzability and traceability.

\section{Related work}\label{sec::RelatedWork}
Control systems evolve from hierarchical, controller-centric structures toward a flatter, network-centric architecture, enhancing interconnectivity and facilitating communication with cloud services and edge devices~\cite{Akerberg2021FutureIndNetw,Leander2023NCC}. These advancements have been leveraged for fault tolerance—employing backup DCNs in the cloud or orchestrators to recover from DCN failures~ \cite{Hegazy2015AutomationAsCloudService,Johansson2022K8sDcsRed}. To our knowledge, the NRP FD algorithm is the first effort to reduce the CAP theorem's~\cite{Gilbert2002CAP} availability tradeoff while preserving consistency in DCN redundancy scenarios~\cite{Johansson2023NRPFD}.
The tradeoff mandated by the CAP theorem is evident in today's redundant DCN systems. Control system user manuals concretize the tradeoff with the different approaches described, which either strive to maintain consistency or prioritize availability upon redundancy link failure~\cite{Siemens2023RedSys,Emerson2022CpuRed}. Fault tolerance is ensured using duplicate links, as depicted in Figure~\ref{fig:SimpRedDCN}. With duplicated links, consistency can be prioritized by disabling DCN redundancy if one link fails~\cite{Siemens2023RedSys}. However, a dual primary situation arises if both links fail simultaneously. Vice versa, availability is prioritized by not disabling redundancy upon one link failure~\cite{Emerson2022CpuRed}. The Leasing NRP FD version assures consistency by maintaining a single primary in all failure scenarios.
 
Appointing a primary is a leader election problem, and various leader election algorithms exist, such as the well-known Bully algorithm~\cite{Garcia-Molina1982bully}. However, the Bully algorithm, and variants thereof, elects multiple leaders in networking partitioning situations, one leader per partition. Alternatively, consensus protocols like Raft and Paxos require a majority~\cite{Ongaro2014RAFT,Lamport2001Paxos}, ensuring consistency even when partitions occur, as only the majority-containing partition progresses. However, the most common DCN redundancy configurations, typically comprising a primary and a backup, do not allow a majority to form in the event of a partition separating the DCNs~\cite{Simion2023PLCRedReview}. The NRP FD method introduces the NRP that, in combination with a DCN, establishes a majority~\cite{Johansson2023NRPFD}. The NRP could be as simple as a layer two network switch responding to an ICMP Ping, providing a means to favor consistency over availability. This paper describes the modeling and verification of the NRP FD strategy, along with a novel, lightweight enhancement ensuring a single primary, i.e., guaranteeing that consistency is preserved due to more than one DCN taking the primary role.
The algorithm is being extended in different directions, considering different configurations and features. Our aim is to  enrich our model align with the extensions of NRP FD, when the extensions are available.

\section{Conclusion and Future Work}\label{sec::Conclusion}
In this paper we describe the process of modeling and formal verification of NRP FD protocol which is used for preserving consistency in DCN redundancy scenarios using Timed Rebeca and Afra. We investigate different failure scenarios and identify situations where network partitioning can lead to a dual primary. We propose an extension, Leasing NRP FD, which preserves consistency and ensures robustness against different failures. 
For future research, we focus on the extensibility and flexibility of the proposed protocol including the exploration of a dynamic network topology, multiple backups and multiple primaries. The latter could be a redundancy plan with a single backup for multiple primaries, each with different and unique characteristics such as specific heartbeat time and network delay. Additionally, we aim to incorporate probability considerations rather than just focusing on the possibility (of failures). 
As another future direction, we plan to investigate the availability trade-off. While NRP-FD prioritizes consistency, this may result in compromising availability. Quantifying this trade-off is a potential direction for further research.

\section*{Acknowledgment}
We acknowledge the support of the Swedish Knowledge Foundation via the synergy project SACSys (Safe and Secure Adaptive Collaborative Systems) and the Profile DPAC (Dependable Platforms for Autonomous Systems and Control). 
We also acknowledge the support of the Swedish Foundation for Strategic Research (SSF) via the Serendipity project.

\bibliographystyle{eptcs}
\bibliography{generic}

\appendix

\section{Rebeca Syntax}\label{app::AbstractSyntaxofTimedRebeca}
\begin{figure}[htb!]
	\centering
    \includegraphics[scale=0.44, bb=0 0 930 370]{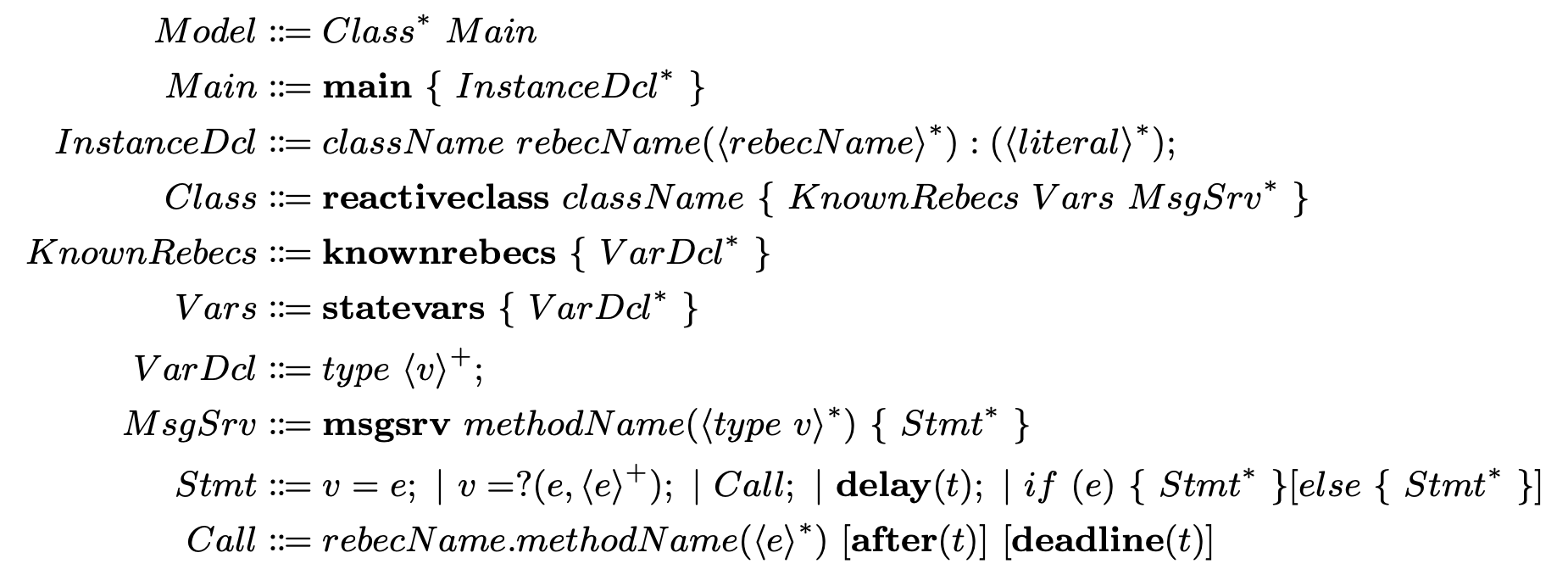}
    \caption{An abstract syntax for Timed Rebeca. 
 The identifiers className, rebecName, methodName, literal and type are self-explanatory. The identifier v denotes a variable. The symbol e denotes an expression, which can be either arithmetic, boolean or a non-deterministic choice. Angular brackets $\langle...\rangle$ serve as meta-parenthesis, with superscript $+$ denoting at least one repetition and superscript $*$ denoting zero or more repetitions. Meanwhile, the use of $\langle...\rangle$ with repetition indicates a comma-separated list. Square brackets $[...]$ indicate that the enclosed text is optional \cite{Sirjani2018Friendliness}. 
   \label{fig:SyntaxTimedRebeca}}  
\end{figure}

An abstract syntax of Timed Rebeca is provided in Figure~\ref{fig:SyntaxTimedRebeca}. 

\newpage
\section{Afra \label{app::Afra}}
\begin{figure}[htb!]
\begin{center}
\includegraphics[scale=0.35, bb=0 0 1230 840]{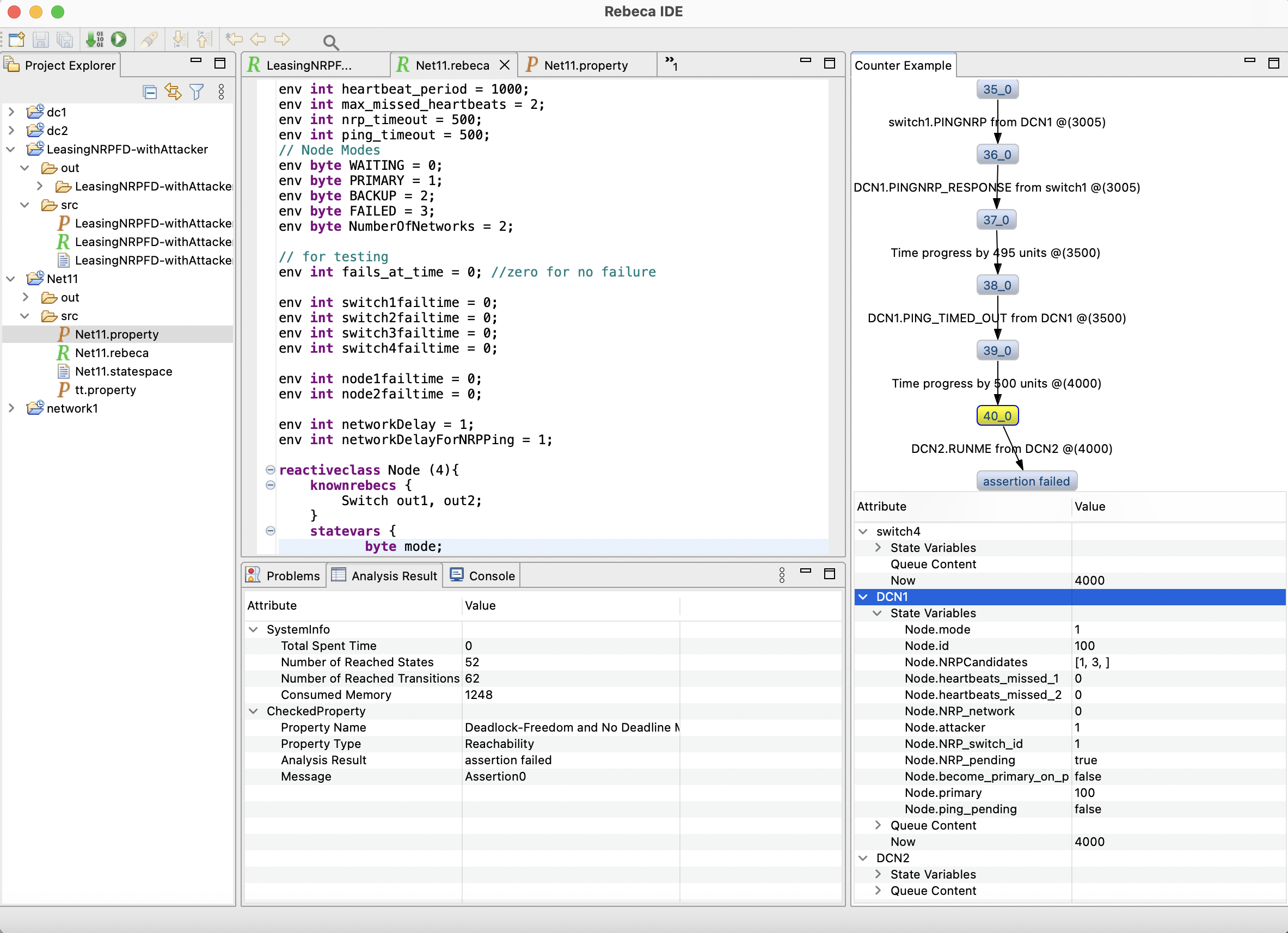}
\end{center}
\caption{A snapshot of Afra.}
\label{fig:Afra}
\end{figure}

A snapshot of Afra is provided in Figure~\ref{fig:Afra}. The state space statistics are shown in the bottom middle.
The generated counterexample is shown in the top right of the panel. At the bottom right, we can see the value of the state variables in each selected state from the counterexample. 

\section{Timed Rebeca model of the Leasing NRP FD
\label{app::TimedRebecaNRPFD}}
In the following the Timed Rebeca model of the Leasing NRP FD is provided. 

\begin{listing}[H]
\scriptsize
\begin{minted}[frame=lines,frame=lines,framesep=1mm,xleftmargin=0em,linenos,escapeinside='',firstnumber=1]{java}
env int heartbeat_period = 1000;
env int max_missed_heartbeats = 2;
env int ping_timeout =100;  
env int nrp_timeout = 100;
// Node Modes
env byte WAITING = 0;
env byte PRIMARY = 1;
env byte BACKUP = 2;
env byte FAILED = 3;
env byte NumberOfNetworks = 2;

env byte MAX_SWITCHES = 99;
// for testing
env int fails_at_time = 0; //zero for no failure

env int switchA1failtime = 0;
env int switchA2failtime = 0;
env int switchA3failtime = 0;
env int switchB1failtime = 0;
env int switchB2failtime = 0;
env int switchB3failtime = 0;

env int node1failtime = 0;
env int node2failtime = 0;

env int networkDelay = 1;
env int networkDelayForNRPPing = 1;

reactiveclass Node (4){
    knownrebecs {
        Switch out1, out2;
    }
    statevars {
        byte mode;
        int id;
        int [2] NRPCandidates;
        int heartbeats_missed_1;
        int heartbeats_missed_2;
        int NRP_network;
        int attacker;
        int which;
        boolean prevWhich;
        int NRP_switch_id;
        boolean NRP_pending;
        boolean become_primary_on_ping_response;  
        int primary;
        boolean ping_pending;
        boolean init;
    }
    Node (int Myid, int Myprimary, int NRPCan1_id, int NRPCan2_id, int myFailTime) {
        id = Myid;
        attacker = 0;
        which=0;
        prevWhich=true;
        NRPCandidates[0] =NRPCan1_id;
        NRPCandidates[1] =NRPCan2_id;	
        heartbeats_missed_1 = 0;
        heartbeats_missed_2 = 0;
        NRP_network = -1;
        NRP_switch_id = -1;
        NRP_pending = true;
        become_primary_on_ping_response = false;
        primary = Myprimary;
        ping_pending = false;
        init=true;
           
        mode = WAITING;
        if(myFailTime!=0) nodeFail() after(myFailTime);
        runMe();
    }
\end{minted}
\label{lst:FullRebecaCode-LeasgNRPFD2}
\end{listing}

\begin{listing}[H]
\scriptsize
\begin{minted}[frame=lines,frame=lines,framesep=1mm,xleftmargin=0em,linenos,escapeinside='',firstnumber=71]{java}
    msgsrv new_NRP_request_timed_out() {
        // if(?(true,false)) nodeFail();
        if (mode == BACKUP) {
            if (NRP_pending) {
                NRP_pending = false;
                if (become_primary_on_ping_response)
                    become_primary_on_ping_response = false;
            }
        }
    }
    // logical action ping_timed_out(ping_timeout)
    msgsrv ping_timed_out() {
        // if(?(true,false)) nodeFail();
        if (mode == BACKUP) {
            if (ping_pending) ping_pending = false;
            else{
                if(which>1){
                    mode = PRIMARY;
                    heartbeats_missed_1 = 0; 
                    heartbeats_missed_2 = 0;
                    primary=id;
                    if(NRP_network==0) out1.new_NRPBack(id, id,NRP_network, NRP_switch_id);
                    else out2.new_NRPBack(id,id, NRP_network, NRP_switch_id);
                    mode = PRIMARY;
                    heartbeats_missed_1 = 0; 
                    heartbeats_missed_2 = 0;
                    primary=id;
                    NRP_pending = true;
                }else NRP_pending = true;
            }
        }else if (mode == PRIMARY){
            if (ping_pending){
                NRP_network++;
                if(NRP_network<NumberOfNetworks){
                    NRP_switch_id = NRPCandidates[NRP_network];
                    if(NRP_network==0) out1.new_NRP(id, id,NRP_network, NRP_switch_id);
                    else out2.new_NRP(id,id, NRP_network, NRP_switch_id);
                } else {
                    NRP_network=NumberOfNetworks;
                    mode=  WAITING;
                }  
                NRP_pending = true;
            } else{
                if(attacker<1){
                    out1.heartBeat(0, id) after(networkDelay);
                    out2.heartBeat(1, id) after(networkDelay);
                }
            }
        }
    }
    msgsrv pingNRP_response(int mid, boolean w, boolean pw){
        // if(?(true,false)) nodeFail();
        if (mode==WAITING); 
        else if (mode == BACKUP){
            if(!w && !pw) which++;
            else which=0;
            if(which>1)
            ping_pending = false;
        }
        else if (mode == PRIMARY) 
            ping_pending = false;
        else if (mode==FAILED);
    }
    msgsrv new_NRP(int mid,int prim, int mNRP_network, int mNRP_switch_id) {
        // if(?(true,false)) nodeFail();
        if(mode!= FAILED){
            NRP_network = mNRP_network;
            NRP_switch_id = mNRP_switch_id;
        }
    }
\end{minted}
\label{lst:FullRebecaCode-LeasgNRPFD3}
\end{listing}

\begin{listing}[H]
\scriptsize
\begin{minted}[frame=lines,frame=lines,framesep=1mm,xleftmargin=0em,linenos,escapeinside='',firstnumber=141,breaklines]{java}
    msgsrv new_NRPBack(int mid,int prim, int mNRP_network, int mNRP_switch_id) {
        // if(?(true,false)) nodeFail();
        if(mode!= FAILED){
            NRP_network = mNRP_network;
            NRP_switch_id = mNRP_switch_id;
        }
    }
    msgsrv runMe(){
      switch(mode){
        case 0: //WAITING : 
            if(init){
                if (id == primary){
                    mode = PRIMARY;
                    NRP_network++;
                    if(NRP_network<NumberOfNetworks){
                        NRP_switch_id = NRPCandidates[NRP_network];
                        if(NRP_network==0)out1.new_NRP(id,id, NRP_network, NRP_switch_id);
                        else out2.new_NRP(id,id, NRP_network, NRP_switch_id);
                    } else NRP_network=NumberOfNetworks;
                } else mode =BACKUP;
                init=false;
            }
            break;
        case 1: //PRIMARY : 
            attacker++;
            if(attacker>1) attacker=1;
            if(NRP_network==0){	
                ping_pending = true;
                out1.pingNRP(id,id, NRP_switch_id) after(5);
                ping_timed_out() after(ping_timeout);
            }else{
                ping_pending = true;
                out2.pingNRP(id,id, NRP_switch_id) after(5);
                ping_timed_out() after(ping_timeout);
            }
            NRP_pending = true;
            break;
        case 2: //BACKUP :
            heartbeats_missed_1++;
            heartbeats_missed_2++;
            if (heartbeats_missed_1 > max_missed_heartbeats && heartbeats_missed_2 > max_missed_heartbeats){
               heartbeats_missed_1 = (heartbeats_missed_1>max_missed_heartbeats+2)?max_missed_heartbeats+2:heartbeats_missed_1;
               heartbeats_missed_2 = (heartbeats_missed_2>max_missed_heartbeats+2)?max_missed_heartbeats+2:heartbeats_missed_2;
                // if(heartbeats_missed_1==heartbeats_missed_2 && heartbeats_missed_2==max_missed_heartbeats+1){
                    // mode = PRIMARY;
                    // heartbeats_missed_1 = 0; // Prevent detecting again immediately.
                    // heartbeats_missed_2 = 0;
                    // primary=id;
                    // NRP_pending = true;
                // }else{
                if(NRP_network==0){
                    ping_pending = true;
                    //NRP_network=-1;
                    out1.pingNRP(id,id, NRP_switch_id) after(15);
                    ping_timed_out() after(ping_timeout);
                }else{
                    ping_pending = true;
                    //NRP_network=-1;
                    out2.pingNRP(id,id, NRP_switch_id) after(15);
                    ping_timed_out() after(ping_timeout);
                }
                NRP_pending = true;
                // }
            }else if(heartbeats_missed_1 > max_missed_heartbeats|| heartbeats_missed_2 > max_missed_heartbeats){
                if(NRP_network==0 && heartbeats_missed_1 > max_missed_heartbeats) {        
\end{minted}
\label{lst:FullRebecaCode-LeasgNRPFD4}
\end{listing}

\begin{listing}[H]
\scriptsize
\begin{minted}[frame=lines,frame=lines,framesep=1mm,xleftmargin=0em,linenos,escapeinside='',firstnumber=209,breaklines]{java}
                    ping_pending = true;
                    out1.pingNRP(id,id, NRP_switch_id) after(5);
                    ping_timed_out() after(ping_timeout);
                }else if(NRP_network==1 && heartbeats_missed_2 > max_missed_heartbeats){
                    ping_pending = true;
                    out2.pingNRP(id,id, NRP_switch_id) after(5);
                    ping_timed_out() after(ping_timeout);
                }
               heartbeats_missed_1 = (heartbeats_missed_1>max_missed_heartbeats+2)?max_missed_heartbeats+2:heartbeats_missed_1;
               heartbeats_missed_2 = (heartbeats_missed_2>max_missed_heartbeats+2)?max_missed_heartbeats+2:heartbeats_missed_2;
            }
            break;
        case 3: //FAILED : 
            break;
      } 
      self.runMe() after(heartbeat_period);
    }
    msgsrv heartBeat(byte networkId, int senderid) {
        // if(?(true,false)) nodeFail();
        if (mode==BACKUP){
            if (networkId == 0) heartbeats_missed_1 = 0;
            else heartbeats_missed_2 = 0;
        }
    }
    msgsrv nodeFail(){
        primary=-1;
        mode = FAILED;
        NRP_network=-1;
        NRP_switch_id=-1;
        heartbeats_missed_1 = 0;
        heartbeats_missed_2 = 0;
        NRP_pending = true;
        become_primary_on_ping_response = false;
        ping_pending = false;
     }
}
reactiveclass Switch(10){
    knownrebecs {
       Node nodeTarget1;
    }
    statevars {
        byte mynetworkId;
        int id;
        boolean which;
        boolean prevWhich;
        boolean failed;
        boolean amINRP;
        boolean primaryPinged;
        boolean terminal;
        Switch switchTarget1;
        Switch switchTarget2;
        int primary;
    }
    Switch (int myid, byte networkId, boolean endSwitch , Switch sw1, Switch sw2, int myFailTime) {
        mynetworkId = networkId;
        primary=0;
        id = myid;
        primaryPinged=false;
        terminal=endSwitch;
        amINRP = false;
        failed = false;
        switchTarget1 = sw1;
        switchTarget2 = sw2;
        which=true;
        if (myFailTime!=0) switchFail() after(myFailTime);
    }
    msgsrv switchFail(){
        failed = true;
\end{minted}
\label{lst:FullRebecaCode-LeasgNRPFD5}
\end{listing}

\begin{listing}[H]
\scriptsize
\begin{minted}[frame=lines,frame=lines,framesep=1mm,xleftmargin=0em,linenos,escapeinside='',firstnumber=276,breaklines]{java}
        amINRP=false;
    }
    msgsrv pingNRP_response(int senderNode,boolean w,boolean pw){
        // if(?(true,false)) switchFail();
        if(!failed)
            if(terminal && senderNode <= MAX_SWITCHES) nodeTarget1.pingNRP_response(id, w,pw); //Pass back
            else if(senderNode >id) switchTarget1.pingNRP_response(id, w,pw);
        else switchTarget2.pingNRP_response(id, w,pw);
    }
    msgsrv pingNRP(int switchNode, int senderNode, int NRP) {
        // if(?(true,false)) switchFail();
        if(!failed)
            if(terminal && NRP==id){
                prevWhich = which;
                which= (senderNode==primary);
                if(switchNode <= MAX_SWITCHES) switchTarget1.pingNRP_response(id,which, prevWhich); //Response
                else nodeTarget1.pingNRP_response(id,which, prevWhich);
            }else if(switchNode >id) switchTarget1.pingNRP(id,senderNode,NRP); 
        else switchTarget2.pingNRP(id,senderNode, NRP);
    }
    msgsrv new_NRP(int senderNode,int prim, int mNRP_network, int mNRP_switch_id) {
        // if(?(true,false)) switchFail();
        if(!failed){
            if(id==mNRP_switch_id) {
                amINRP=true;
                primary=prim;
            } else amINRP=false;
            if(terminal && senderNode <= MAX_SWITCHES)nodeTarget1.new_NRP(id,prim, mNRP_network, mNRP_switch_id);
            else if(senderNode >id) switchTarget1.new_NRP(id,prim, mNRP_network, mNRP_switch_id); //Pass back
            else switchTarget2.new_NRP(id,prim, mNRP_network, mNRP_switch_id);
        }
    }
    msgsrv new_NRPBack(int senderNode,int prim, int mNRP_network, int mNRP_switch_id) {
        // if(?(true,false)) switchFail();
        if(!failed){
            if(id==mNRP_switch_id) {
                amINRP=true;
                primary=prim;
            } else amINRP=false;
            if(terminal && senderNode <= MAX_SWITCHES)nodeTarget1.new_NRPBack(id,prim, mNRP_network, mNRP_switch_id);
            else if(senderNode >id) switchTarget1.new_NRPBack(id,prim, mNRP_network, mNRP_switch_id); //Pass back
                else switchTarget2.new_NRPBack(id,prim, mNRP_network, mNRP_switch_id);
        }
    }
    msgsrv heartBeat(byte networkId, int senderNode) {
        // if(?(true,false)) switchFail();
        if(!failed)
            if(terminal && senderNode <= MAX_SWITCHES) nodeTarget1.heartBeat(networkId,id) after(networkDelay);
            else if(senderNode > id) switchTarget1.heartBeat(networkId,id) after(networkDelay);
                else switchTarget2.heartBeat(networkId,id) after(networkDelay);
    }
}

main {
    @Priority(1) Switch switchA1(DCN1):(1, 0, true , switchA2 , switchA2 , switchA1failtime);
    @Priority(1) Switch switchA2(DCN1):(2 ,0, false , switchA1 , switchA3 , switchA1failtime);
    @Priority(1) Switch switchA3(DCN2):(3, 0, true , switchA2 , switchA2 , switchA3failtime);
    @Priority(1) Switch switchB1(DCN1):(4, 1, true , switchB2 , switchB2 , switchB1failtime);
    @Priority(1) Switch switchB2(DCN1):(5, 1, false , switchB1 , switchB3 , switchB1failtime);
    @Priority(1) Switch switchB3(DCN2):(6, 1, true , switchB2 , switchB2 , switchB3failtime);
    
    @Priority(2) Node DCN1(switchA1, switchB1):(100, 100, 1, 4, node1failtime);
    @Priority(2) Node DCN2(switchA3, switchB3):(101, 100, 3, 6, node2failtime);
}
\end{minted}
\label{lst:FullRebecaCode-LeasgNRPFD6}
\end{listing}

\newpage
\section{State Space \label{app::StateSpace}}
The state space of Timed Rebeca model for the  NRP FD (including the problematic optimization) implementing case 7 of Table \ref{tab:configurationofFailures} has 70 states and 88 transitions.
 Case 7 is where switchA1 and switchB1 fail simultaneously at time 2500. 
A portion of the visualized state space is provided in Figure~\ref{fig:StateSpace-NRPFD}.  
We define the followings in the the property file (see L\ref{Property}): 
    \vspace{-2mm}
    \lstset{language=Java}
    \begin{lstlisting}[belowskip=-0.2 \baselineskip] 
    DCN1Primary = (DCN1.mode ==1);
    DCN2Primary = (DCN2.mode ==1);
    DCN2Backup  = (DCN2.mode ==2);  
    switchA1Failed  = (switchA1.failed);
    switchB1Failed  = (switchB1.failed);
    switchA1NRP  = (DCN1.NRP_switch_id==1 && DCN2.NRP_switch_id==1);
    ...
    \end{lstlisting}
    \vspace{2mm}
The term $DCN1Primary$ means that the mode of DCN1 is PRIMARY (similar for DCN2) and the term $switchA1Failed$ 
means that the state variable $failed$ of switcheA1 is $true$ (similar for switcheB1). $switchA1NRP$ means that the state variable $NRP\_switch\_id$  equals 1 (the id of switchA1) for both DCNs. 
In case 7 of Table \ref{tab:configurationofFailures}, both switches fail at time 2500. As we are at the time 3000 in S59, switchA1Failed and switchB1Failed are true at the states depicted.  
Both DCN1 and DCN2 execute a runMe in each heartbeat period: 
    \vspace{-2mm}
    \lstset{language=Java}
    \begin{lstlisting}[belowskip=-0.2 \baselineskip] 
    heartbeat_period = 1000 // line 1 of Listing 1
    ...
    self.runMe() after(heartbeat_period) // line 35 of Listing 1
    ..
    \end{lstlisting}
    \vspace{2mm}

In each period, PRIMARY (DCN1) checks its NRP availability. In the state S63, DCN1 sends a PINGNRP message to switchA1 in the new heartbeat period, @3000. By receiving PINGNRP, switchA1 which is failed, does nothing (line 296 of Appendix C).
In the state S65, by running Ping\_timed\_out, DCN1 will notice that switchA1 has failed. DCN1 tries to select a new NRP from its NRP candidate set (here switchB1 which is not operational at the moment). Note that there is no active NRP in S66. 
At the next runMe, @4000, DCN2 changes its mode to PRIMARY due to missing 
more than maximum heartbeats allowed on both networks simultaneously.
We can see in S70  a dual primary situation occurred. 
We commented out the assertion such that the model checker continues creating the state space.


\begin{figure}[htp]
\centering
\begin{tikzpicture}[spy using outlines={circle,yellow,magnification=2.5,size=5.5cm, connect spies}]
\node {
\includegraphics[scale=0.19, bb=0 0 530 2300]{Case7-Partial.pdf}};
\spy on (0,2.3) in node [left] at (8,5);
\spy on (0.2,-1.7) in node [left] at (10,-0.25);
\spy on (0.2,-5.3) in node [left] at (8,-5.5);
\end{tikzpicture}
\caption{A part of the visualized state space for the Timed Rebeca model of the NRP FD.}
\label{fig:StateSpace-NRPFD}
\end{figure}

\end{document}